\documentclass[12pt,preprint,aps,nofootinbib,showkeys,amsmath,amssymb]{revtex4}
\usepackage[dvips]{graphicx}
\usepackage{subfigure}

\newcommand {\be}{\begin{equation}}
\newcommand {\ee}{\end{equation}}

\begin{document}
\title{Coupling different levels of resolution in molecular simulations}
\author{Sim\'on Poblete}
\author{Matej Praprotnik}
\altaffiliation{On leave from the National Institute of Chemistry, Hajdrihova 19, SI-1001 Ljubljana, Slovenia}
\author{Kurt Kremer}
\email{kremer@mpip-mainz.mpg.de}
\author{Luigi Delle Site}
\email{dellsite@mpip-mainz.mpg.de}
\affiliation{Max-Planck-Institut f\"ur Polymerforschung, Ackermannweg 10, D-55128 Mainz, Germany}

\begin{abstract}
Simulation schemes that allow to change 
molecular representation in a subvolume of the simulation box
while preserving the equilibrium with the surrounding
introduce conceptual problems of thermodynamic consistency.
In this work we present a general scheme based on thermodynamic arguments which ensures thermodynamic equilibrium among the molecules of different
representation. The robustness of the algorithm is tested for two 
examples, namely an adaptive resolution simulation, atomistic/coarse-grained, for a liquid
of tetrahedral molecules and an adaptive resolution simulation of a
binary mixture of tetrahedral molecules and spherical solutes.

\end{abstract}

\maketitle
\newpage
\section{Introduction}
In many complex systems, rather small local changes,
e.g. of molecular structure due to a mutation or complexation of
charged molecular fragments due to the presence of multivalent
salt, can have significant consequences for their global
properties and function. To understand such phenomena and then
derive construction or processing principles from the derived
knowledge is one of the key areas of modern basic materials
science and related areas. To achieve that goal analytic theory
and experiment can only provide limited input. Analytic theory can
only deal with highly idealized limiting cases and thus provide
rigorous anchor points for any other treatment. On the other hand,
experimentally it is most often not possible to provide a detailed
microscopic characterization down to the atomistic level in all
necessary detail. Because of this numerical modeling is more and
more becoming an indispensable tool. The questions described pose
central challenges for computational studies. It is in most cases
impossible to perform simulations of systems such as a polymer melt, a
biological systems of proteins in a membrane or a solution of
larger molecules in water on an atomistically detailed level for a
long enough time (assuming that atomistic force fields of reliable
quality exist). Because of that a variety of multiscale simulation
techniques, ranging from straightforward hierarchical
parameterizations, see e.g. \cite{prllu,jacs1,jacs2,
karsten1,karsten2,florianjcp,vothbook}, to interfaced layers of
different resolutions see e.g.
\cite{kax,Rottler:2002,Csanyi:2004,Laio:2002,Jiang:2004,Lu:2005} have been developed.
Even if one could perform the all atom simulations mentioned
above, the reduction of the huge amount of data to extract an
understanding of phenomena and mechanisms already requires a
systematic coarse graining. While most techniques are sequential
in a way that at a given time the whole system is described with
the same representation (level of resolution), in many cases it
would be convenient to locally adjust on-the-fly the level of
resolution according to the properties of interest, while keeping
the larger surrounding on a coarser level. Such an approach would
however require full equilibrium between these different regimes.

A typical example is the solvation of a molecule in water where
the interesting physics and chemistry occurs within few solvation
shells around the molecule while outside the water plays the role
of a  source of mesoscopic water molecules. Thus one has to
interface different molecular models of water (e.g. flexible,
rigid, and coarse-grained) and to allow for the exchange of
molecules among the different representations. This would properly
characterize the relevant physics and chemistry of each region and
would properly describe the local fluctuations. The concept above
can be generalized in terms of designing an algorithm which
interfaces two different force-fields describing the same
molecules where the exchange of particles from one region of
representation to another (and vice versa) occurs under
equilibrium conditions. If this can be done on the basis of a
rather general framework, it would allow also to couple rather
loosely connected representations, as will be discussed below.

Such force fields may have the same level of resolution (i.e. the
molecule carries the same number of degrees of freedom, (DOFs)) or
different resolution (for example atomistic and coarse-grained).
To the first class of problems belong approaches as those of {\it
learn on-the-fly} \cite{Csanyi:2004} where in a certain region of
space the force acting on the atoms is updated on the fly by
underlying quantum calculations and is interfaced with a standard
classical force-field which describes the interactions in the
larger region outside. The number of DOFs may remain the same but
the force acting on each atom are different in the different
regions. To the second kind of problems belong approaches as those
of adaptive resolutions \cite{jcp,pre1,annurev,ensing,heyden}
where the molecular models carry different number of DOFs.
Extensions to link such particle based approaches to
continuum have also successfully been tested\cite{delgado:2008,delgado:2009}. 
So far, a
physically consistent theoretically framework which properly
describes the change of representation and automatically leads to
thermodynamic equilibrium between the different regimes is still
missing. Here we study this problem and propose a general
framework for such a kind of approaches.

\section{Adaptive Resolution Concepts}
The underlying idea for going from one molecular representation to
another is to introduce a transition region at the interface,
where the molecules slowly change their representation. In this
region they are in equilibrium with their actual surrounding and
change continuously until the region of the new representation is
reached. There they "arrive" fully equilibrated within the
surrounding described by the new representation. At a first
glance a natural way to proceed would be an Energy-based approach
where a smooth space dependent function would interpolate between
the Hamiltonians corresponding to two force-fields. This approach
has been shown to lead to unphysical artifacts inconsistencies
\cite{pre2,jpa,prelu}. To avoid this, we proposed a force based
simulation approach called AdResS (Adaptive Resolution Scheme)
\cite{annurev}. Here we extend this with a thermodynamically
consistent description of the transition regime, which eventually
allows to couple adaptively rather different systems and provides
first step to open systems molecular dynamics simulations.

\begin{figure}[H]
\includegraphics[width=0.5\textwidth]{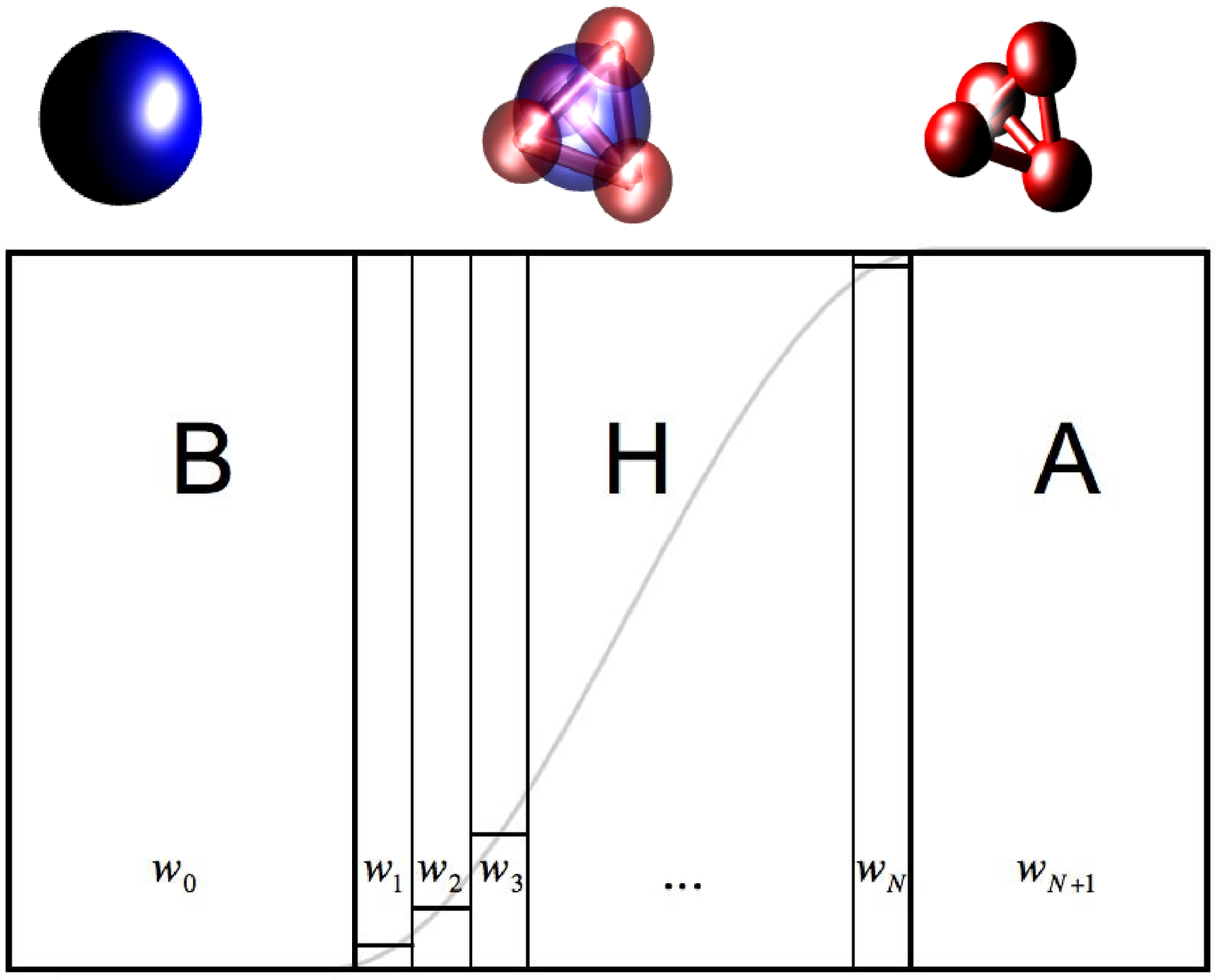}
\caption{Pictorial representation of the adaptive box and molecular representation. The region on the left, indicated by $B$, is the low resolution (coarse grained) region, the central part is the transition (hybrid) region $H$, where the switching function $w(x)$ (curve in grey) is defined, and the region on the right, indicated by $A$ is the high resolution (atomistic) region. For numerical convenience, as discussed in the text, in order to calculate the chemical potential of each resolution, the transition region is divided in $N$ slides which corresponds to discretized values of the switching function, here indicated with $w_{i}$.\label{box}}
\end{figure}

The basic idea is to allow the molecules to experience a smooth
transition from one force-field to the other and vice versa
without altering the equilibrium of the system. For this we
introduce a transition region, where an interpolation function is
defined in terms of the position of the center of mass of a
molecule. As an example, as applied in the AdResS scheme, for a
pair force between molecules $\alpha$ and $\beta$ the formula may
be written as:
\begin{equation}
{\bf F}_{\alpha \beta}=w(X_\alpha)w(X_\beta){\bf
  F}_{\alpha\beta}^{A}+[1-w(X_\alpha)w(X_\beta)]{\bf F}^{B}_{\alpha\beta}
\label{forces}
\end{equation}
where ${\bf F}_{\alpha\beta}^{A}$ is the force obtained from the
potential of representation $A$ and ${\bf F}_{\alpha\beta}^{B}$
the one obtained from the potential of representation $B$; $w(X)$
is the switching function and depends on the center of mass
positions $X_{\alpha}$ and $X_{\beta}$ the two interacting
molecules, as indicated in grey in Fig.\ref{box}.
While with the above approach one can perform an MD simulation and
control the molecular dynamics, the forces as given in Eq.1 cannot
be expressed as the derivative of a Hamiltonian. This rises the
question of how to assure the thermodynamic equilibrium in such a
force based approach. Indeed our previous studies displayed
density fluctuations in the transition regime, which in some cases
had to be repaired by a pressure correction term \cite{pre1}. The
main problem in changing representation in a continuous way is
that the DOFs for which the interaction
becomes different or which are switched on or off in going from
one representation to another are characterized by different
energy functions and thus contribute differently to the global
equilibrium of the system. This process is associated with the
acquisition and release of thermal and interaction energies of
these DOFs which must be slowly redistributed as the new
representation is acquired. Note that the total energy of the
molecules in the different regimes does not at all have to be the
same - in most cases it actually will not be the same! These
energies related to such a process can be viewed as some sort of
latent heat that takes care of the equilibration of the molecules
with their environment (see e.g. \cite{pre2}). Switching on and
off degrees DOFs can be shown to correspond to fractional DOFs and
the related equipartition theorem thus allows to define a
temperature and thus a thermostat in the transition
regime\cite{pre2,jpa}.

\section{Generalized Coupling Scheme and Thermodynamic Driving Force}
The above intuitive ansatz, can be formalized and generalized
within a thermodynamically consistent framework. This theoretical
framework allows to explicitly define equations of motion in the
transition regime by which both the dynamics and the
thermodynamics can be controlled, despite the fact that there is
not a well defined energy as in standard simulation schemes. To do
this we reformulate the problem in specific terms of an additional
thermodynamic force and the internal energy of a molecule as follows.

For the more detailed region $A$ we want to
study it as a  subsystem at a given temperature $T$ in a fixed volume
$V$ with a well defined average number of particles $N$ and
pressure $P$. This has to be coupled to a more coarse grained
surrounding in a way that the structural and dynamical properties
within the region $A$ are (ideally) not altered at all. This also
requires that there is no kinetic barrier introduced by the
transition regime between $A$ and $B$. Viewing the different
regimes as different phases, the question of equilibrium between
different regimes generally can be formulated in terms of the
differences in the chemical potential\cite{note2} characterizing each resolution. To do this let us
consider the difference $\phi(x)=\mu_{A}-\mu_{w(x)}$ between the
chemical potential of a molecule in region $A$ (chemical potential of a system composed solely by high resolution molecules; $w(x)=1$), and in
a hybrid system exclusively composed of hybrid molecules with a fixed level of resolution $0\le w=w(x)=const.\le 1$ corresponding to a fixed bulk value $\mu_{w(x)}$. 
Since $w$ in such a hybrid system is constant within the whole system, 
we now have a well defined
energy function, which allows to determine $\mu_{w(x)}$. 
By repeating this procedure for each value of $x$, we
can now approximate an effective, position dependent, chemical potential of the molecules
for the whole system, especially in the transition regime in
the full range $0\leq w \leq1$. Since in the adaptive
representation scheme $w=w(x)$, $\phi$ becomes a position
dependent function in such a simulation\cite{note3}. 
The wider the transition regime the better this approximation is
expected to work, since the difference in the interactions in the
direction of growing and shrinking $w$ vanishes. This idea can now
be used to couple within one simulation box two systems, where the
same molecules are described by different sets of DOFs. Coupling
two systems along Eq.1 and running the simulation with a regular
Langevin or DPD thermostat\cite{annurev,Junghans} often leads to
the problem of a nonuniform free energy density throughout the
simulation cell, since the free energy density, which to a first
approximation depends on the DOFs per molecule, might be different
in the different regions. This results then in unwanted density
variations especially in the transition regime. As we will see
below, the thermostat generally only compensates for part of this
problem. In this context $\phi(x)$ is nothing else than the
quantity which reintroduces, in an effective way, a formal
uniformity.

To calculate $\phi(x)$, we can divide it into two components. The
first part is due to the potential of interaction (called ''excess
chemical potential''in the following) between the degrees of
freedom, which are switched on or off. The second corresponds to
the kinetic intramolecular part (internal vibrations and molecular
rotations). The latter part typically can also be taken care of by
the thermostat (see below). The calculation of the first component
can be numerically achieved as illustrated in Fig.~\ref{box}.
The simulation box is divided into a region of force-fields $A$
and $B$ and a transition region $H$ in between. The region
$B$ is characterized by the value of the switching function
$w_{0}=0$. The region $A$ is characterized by the value of the
switching function $w_{N+1}=1$. In $H$ the value of $w$ in
the actual simulations varies continuously. However here we
approximate this by discretizing $w$ into N steps $w_{1}$,
$w_{2}$,... $w_{N-1}$, $w_{N}$. For any fixed value of $w$ the
energy function is well defined and the excess chemical potential
then is defined as:
$\mu^{exc}(x_{i})=\mu^{exc}_{w_{i}}$, where the
$\mu^{exc}_{w_{i}}$ is the chemical potential of the molecules in
a bulk system of the specific representation of $w_{i}$. To
calculate numerically each $\mu^{exc}(w_{i})$ one can use standard
particle insertion methods \cite{frenkel}. Repeating this
procedure with all values of $w_{i}$ leads to a position dependent
excess chemical potential $\mu^{exc}(x)$. The implicit
approximation that each of the stripes in Fig.\ref{box} can be
taken as a bulk system and is statistically independent of each
other will be shown to be of minor importance for practical
applications. 
The second component is the ideal gas kinetic
contribution to the chemical potential coming from the internal
degrees of freedom. Usually in a ``one-representation''
simulation, this contribution to the chemical potential is ignored
being only a trivial constant. In our case where the DOFs of
interest might continuously change in going from one
representation to another, each DOF in the transition region
contributes differently according to the corresponding value of
$w(x)$. While such contributions can be easily calculated for the
force-fields $B$ and $A$, the critical aspect to address is what
happens for the hybrid representation in the transition region.

For the AdResS scheme we have shown that the interpretation of
changing representation as continuous change in dimensionality
(between zero and one and vice versa) of the associated phase
space of a DOF (see e.g. \cite{pre2,jpa,annurev}) allows for a
proper definition of the temperature and thus of a thermostat.
This means that if a DOF remains unchanged from one representation
to another, its dimensionality is one (invariant full contribution
to statistical properties regardless of the representation). If
instead a DOF is switched on/off from one representation to
another its dimensionality goes from one to zero or vice versa;
that is if a DOF is explicitly present in force-field $A$ and not
present in force-field $B$ it would not contribute to the
statistical properties in the region $B$, and would gradually
contribute in the transition region up to full contribution in
region $A$. In other words the dimensionality of the phase space
associated with a DOF reflects the degree of ''representation''
expressed by $w(x)$ and it weights, accordingly, its contribution
to the average properties of the system. The formalism of
fractional calculus has been shown to be able to formally describe
this process so that one can calculate the kinetic energy
contribution to the free energy per particle \cite{jpa}. This
means that one can calculate the chemical potential for a given
representation $w$ analytically and, since $w=w(x)$, obtain the
ideal gas contribution. For a generic switchable degree of freedom
$p$ this is written as:
\begin{equation}
 A_{p}=-\beta\log\left[\int e^{-\beta p^{2}}d^{w}p\right]=\mu^{kin}_p(w)
\label{idealeq}
\end{equation}
and thus the total contribution of the entire set of switchable
degrees of freedom (assuming that they decouple) is:
\begin{equation}
 \mu^{kin}(w)=\sum_{DOF}\mu^{kin}_p(w) .
\end{equation}
and the component to the latent heat
$\phi(x)^{kin}=\mu^{kin}_{A}-\mu^{kin} (w)$. The solution of
Eq.\ref{idealeq} can be obtained analytically:
\begin{equation}
\mu^{kin}_{p}(w)=C kT
\left(\frac{w}{2}\right)\log(T)+kT\log\frac{\Gamma\left(\frac{w}{2}\right)}{\Gamma\left({w}\right)}
\label{solideal}
\end{equation}
where $C$ is a
constant, , $k$ the Boltzmann constant, $T$ is the temperature and $\Gamma$ the standard
$\Gamma$ function. The first term in Eq. \ref{solideal} is linear in $w$ and therefore linearly interpolates between the coarse-grained and all-atom values of $\mu^{kin}$\cite{annurev}. Note that the second nonlinear 
term is  negligible in the temperature regime of interest.
At this point we have a numerical definition of
$\mu^{exc}(x)$ and an analytical definition $\mu^{kin}(x)$. 
In general, the idea of continuous interpolation and the calculation of the latent heat presented here have some formal similarities with the method to calculate entropies or chemical potentials, which can
not be calculated directly. This is done by adiabatically coupling via a continuous parameter the real potential to
one with a potential where the chemical potential is known and then integrating
over that parameter \cite{alder}.
For the pourposes of this work $\mu^{exc}(x)$ $\mu^{kin}(x)$ and are all the ingredients to control the thermodynamic equilibrium
of our system and we can use these quantities within the numerical
procedure to couple the different resolutions. $\mu^{exc}(x)$
characterizes the chemical potential due to the interactions among
molecules according to their representations. Thus the gradient of
$\mu^{exc}(x)$ can be interpreted as a thermodynamic force $F_x^{TD}=-\frac{\partial \mu^{exc}}{\partial x}$.
Subtracted from the standard AdResS forces, Eq.\ref{forces}, 
should compensate any drift
originating from the different resolutions, making the density
profile uniform throughout the whole simulation
box\cite{note4}. Similar expressions also 
emerge in interspecies forces in dense binary systems\cite{Zhang}.
Next, $\mu^{kin}$, is the ''internal'' energy of a
molecule, independent from the direct interaction with its
surroundings, and thus $\phi^{kin}$ it is nothing else than the internal heat
that is acquired or removed as the representation changes. Since
the temperature is well defined, this can be supplied by any
standard local thermostat. One can check {\it a posteriori}
that indeed this internal heat is in average exactly the amount
provided by the thermostat. We have applied the concept described
above to two examples, (a) to the adaptive
simulation, atomistic/coarse-grained, for a liquid of tetrahedral
molecules (see Fig.\ref{box} (top)) and (b) to the adaptive simulation of
a binary mixture with major component tetrahedral molecules and with
spherical solutes (see Fig.\ref{cartoon}).

\section{Applications to model systems}
\subsection{(a) Liquid of Tetrahedral Molecules}
\begin{figure}
\centering \mbox{\subfigure
{\includegraphics[angle=0,width=5cm]{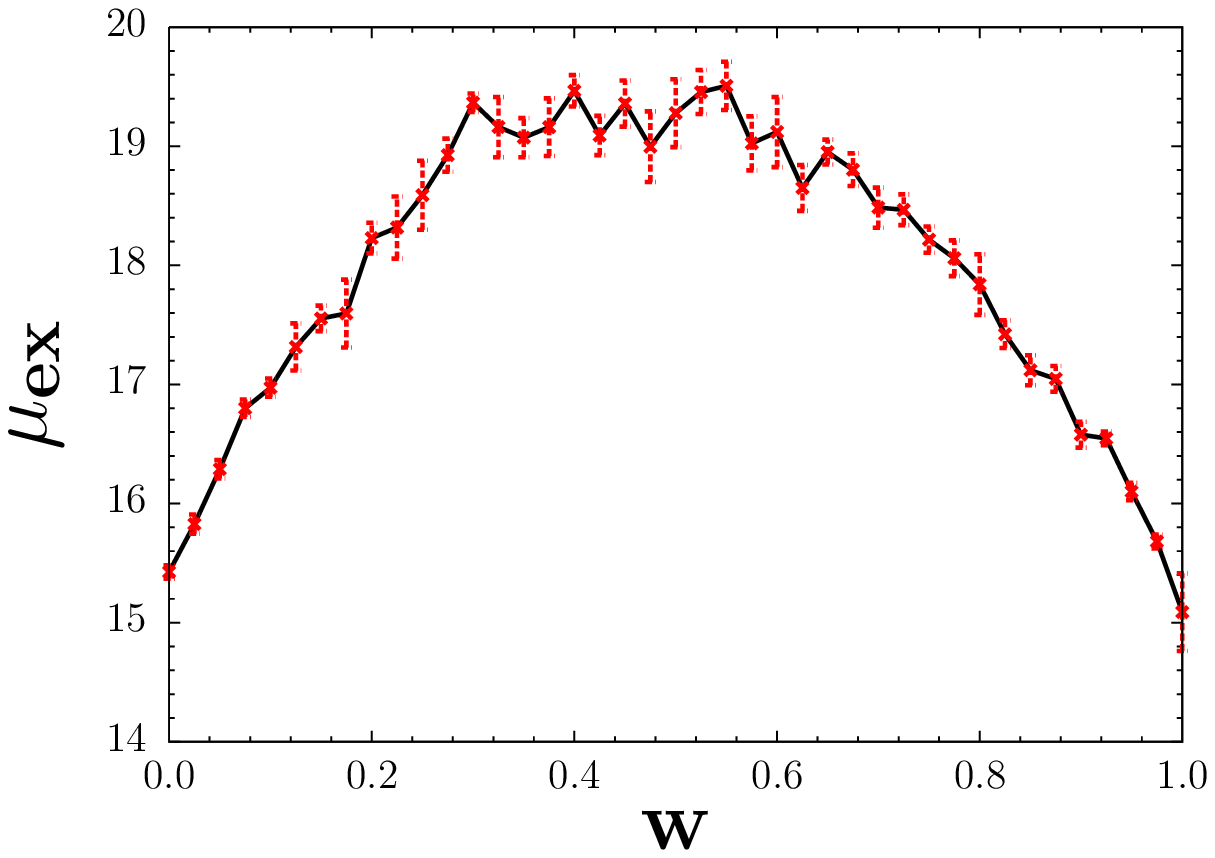}
           \subfigure
           {\includegraphics[width=5cm]{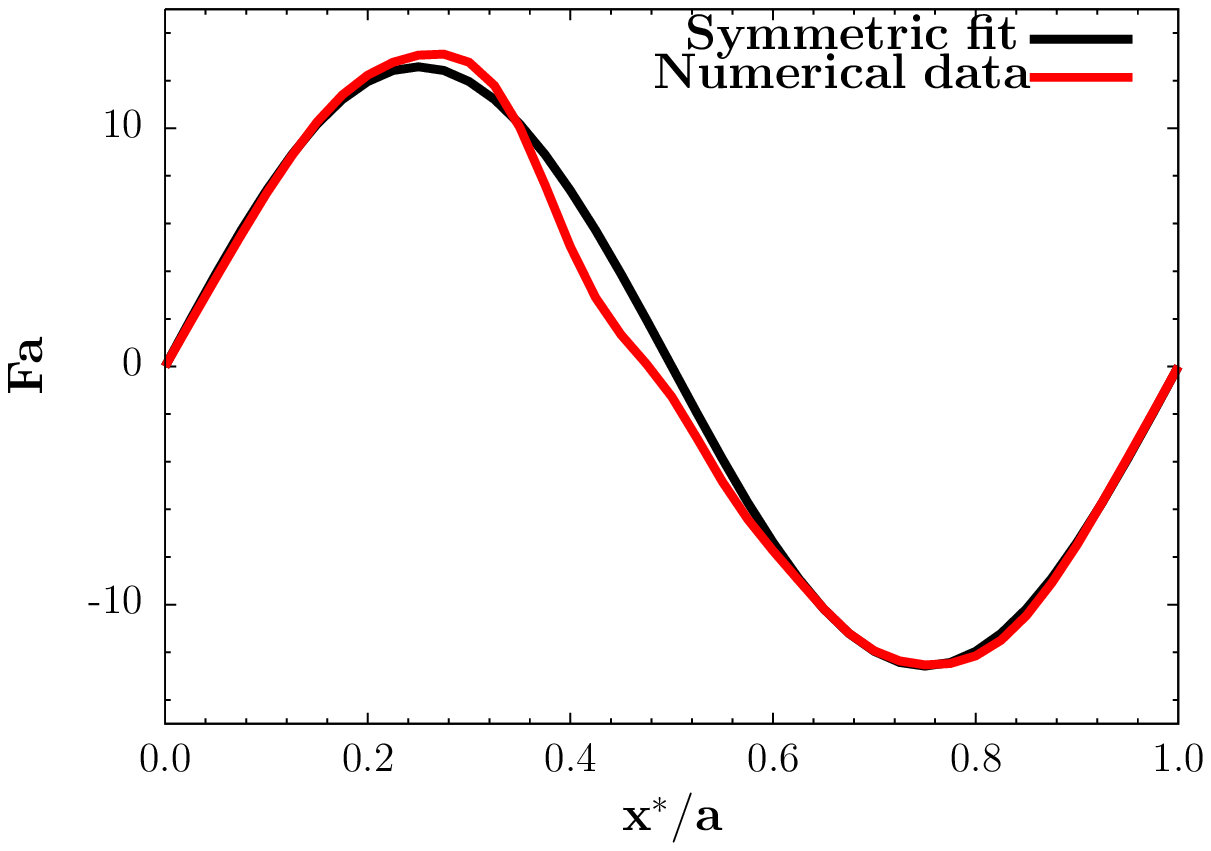}}}}
\caption{Plot of the excess chemical
potential (left) and its thermodynamic force (right). The plot of the force shows, in grey (red) the curve obtained from the numerical data to which a smoothing process was applied and in black the curve obtained from a symmetric fit of the data; both give the same results within the error bar. $a$ is the width of the transition region in reduced units. Because the atomistic and the coarse grained representation describe the same state point they are characterized by the same chemical potential, while this does not hold anymore for the hybrid.
 \label{exchem}}
\end{figure}

\begin{figure}[H]
\includegraphics[width=0.5\textwidth]{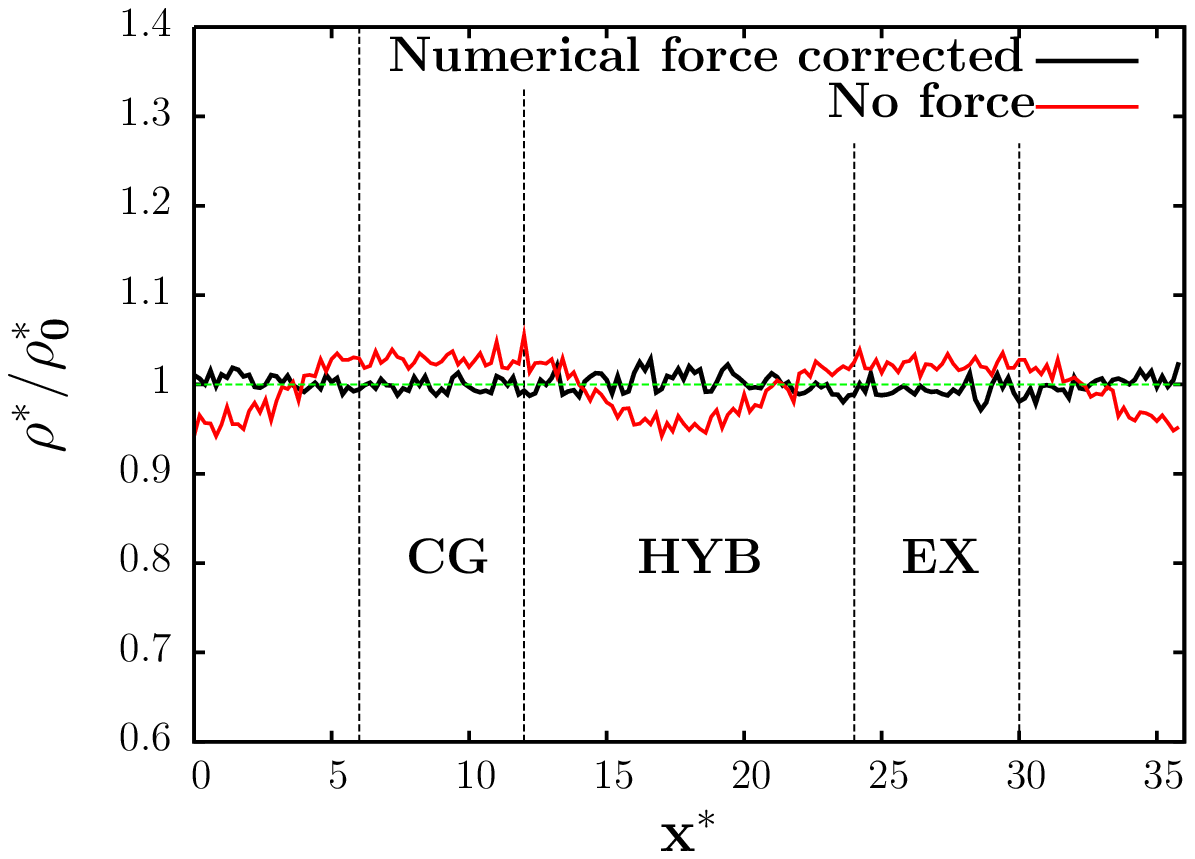}
\caption{Plot of the density across the simulation box. In grey (red) the case without the thermodynamic force, in black the case where the thermodynamic force is applied. In both cases the system is coupled to a locally acting thermostat. The thermodynamic force clearly improves the quality of the results.\label{resgood}}
\end{figure}

We first test the above idea on the  example of simple tetrahedral
molecules, where the molecules can change representation from an atomistic to a coarse-grained resolution and vice versa passing through a series of hybrid representations (see Ref.\cite{jcp} and Figure \ref{box}).
This model system also has been used during the first introduction
of the AdResS method. In such a system an atomistic representation
is interfaced with its corresponding coarse-grained one.
We treat the system at temperature $T=\varepsilon/k$ 
and a liquid density with an atom
density of $\rho=0.175/\sigma^3\approx
1.0/\sigma_{cg}^3$ ($\sigma_{cg}$ is the excluded volume diameter
of the coarse-grained molecule).  Here $\sigma$ and $\varepsilon$
are the standard Lennard-Jones parameters of length and energy, respectively.
For the force field parameters and 
other modeling details see refs.\cite{jcp,pre1}.
The system is set up in such a way,
that the equation of state is the same in both the coarse grained
and in the all atom regime at the temperature and density of the
current simulation. Because of that $\mu^{exc}(x)$ has to be the same
for $w=1$ and for $w=0$. As was found in earlier
applications of the AdResS algorithm, the coarse grained and the
detailed regime are in equilibrium with each other and the
molecules are free to move from one regime into the other while
simultaneously changing their molecular representation. A typical
problem however, as also found in the application of this method
to water, were significant density variations within the hybrid
regime. Employing the above derived scheme and introducing the
corresponding thermodynamic force this problem can be solved.
Fig.\ref{exchem} shows $\mu^{exc}$ and the resulting thermodynamic
force.
The results of the application of such a force plus the local thermostat
for the internal heat is shown in Fig.\ref{resgood} in comparison
to the case without that correction.
As one can clearly see indeed this procedure leads to a more
satisfying density profile which automatically emerges from the
forces applied. Remaining very small deviations from the ideally
flat profile, which are expected due to the rather approximate way
to determine the thermodynamic force, can easily be eliminated in
a short iterative procedure, which optimizes the force. In order
to prove the full consistency of the method we must still show
that the molecular internal heat provided by the thermostat
corresponds to that calculated analytically. To do so, we
calculate the work done by the thermostat as in
Ref.\cite{munakata}. By removing the contributions of the center
of mass we indeed have only the energy corresponding to the
internal degrees of freedom.
\begin{figure}[H]
\includegraphics[width=0.5\textwidth]{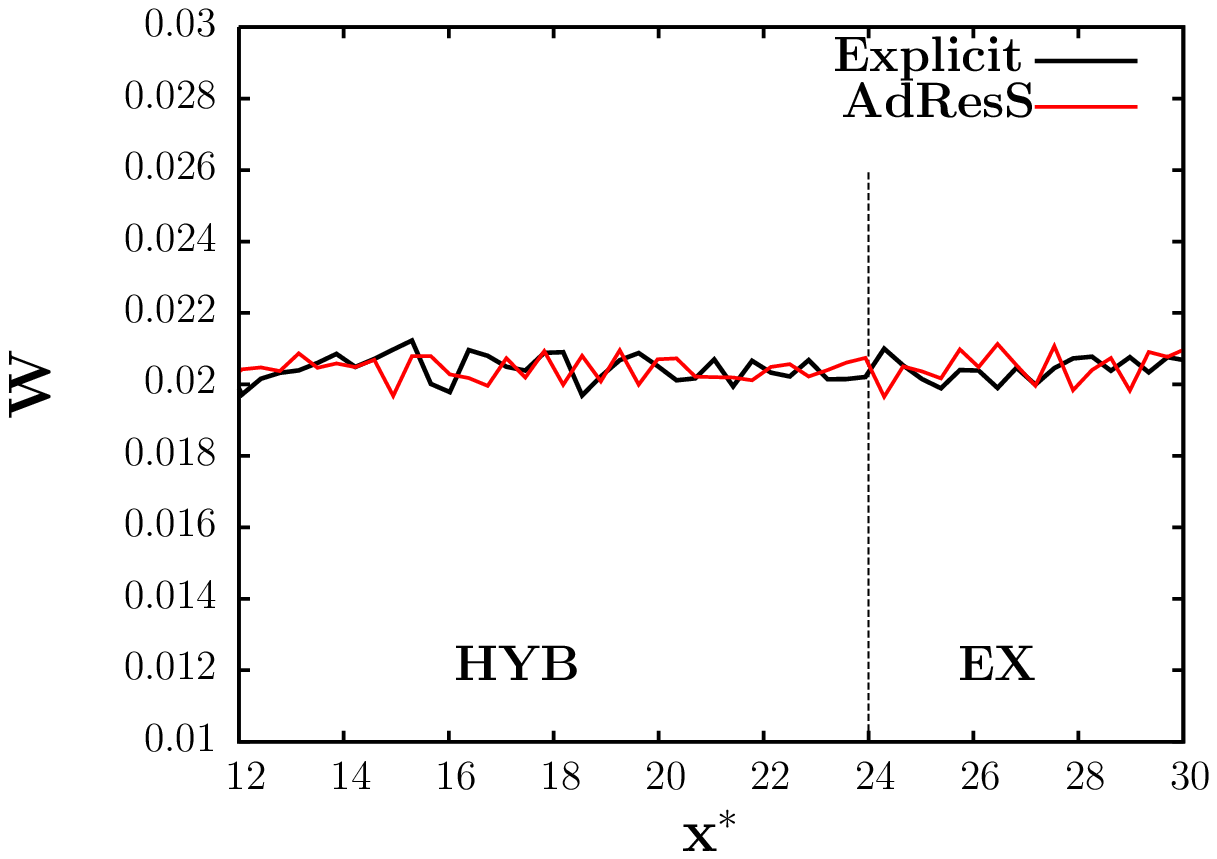}
\caption{Heat profile provided by the thermostat across the box. The heat is calculated  as the work done by the thermostat by removing the contributions of the center of mass. This is the energy corresponding to the
internal degrees of freedom.\label{heatkost}}
\end{figure}
According to the fractional formalism, the explicit average
kinetic energy of a ''switchable'' DOF is $w \cdot<K_{atom}>$
\cite{pre2,jpa}. If a thermostat provides an amount of heat
$<W_{atom}>$ for the atomistic resolution, it provides an explicit
amount of heat $w \cdot<W_{atom}>$ to the hybrid resolutions
\cite{note}. Hence the average extra (latent) heat that the
thermostat effectively provides to a molecule in the transition
region, in order to have the same internal energy as a molecule in
the atomistic resolution, is $<W_{atom}>-w \cdot
<W_{atom}>=(1-w)\cdot<W_{atom}>$, that is proportional to the
first term in the analytical expression of $\phi^{kin}$. This
means that in practice the heat given by the thermostat to the
internal DOF's in the hybrid region is the same as in the
atomistic one as consistently obtained in our calculations and
shown in Fig.\ref{heatkost}, while it counts to the total energy
of the system only according to the value $w$ in of the actual
local resolution. At a first glance the conclusion above seems
obvious and can easily be seen for decoupled DOFs. However,
the coupling between the intra and the intermolecular interactions
is different according to the different resolutions across the
box.
The equation above provides the first order approximation of heat
that must be given by the thermostat according to the formalism
introduced in order to have equilibrium across the whole box. In
this perspective the numerical calculations indeed indicate that
the equation above holds. More important, this proves the
robustness of the algorithm regarding the hypothesis of separation
between intra and intermolecular DOFs; thus it validates the whole
theoretical framework from which the equations governing the
switching are obtained. So far, this example shows the validity of
the idea of thermodynamic force for a one component system only.
However, to apply such an approach to more interesting problems
from biophysics or physical chemistry and material science, an
extension to the case of multi component systems such as mixtures
is needed.

\subsection{Adaptive Resolution Simulation of a Mixture}
\begin{figure}[H]
\includegraphics[width=0.5\textwidth]{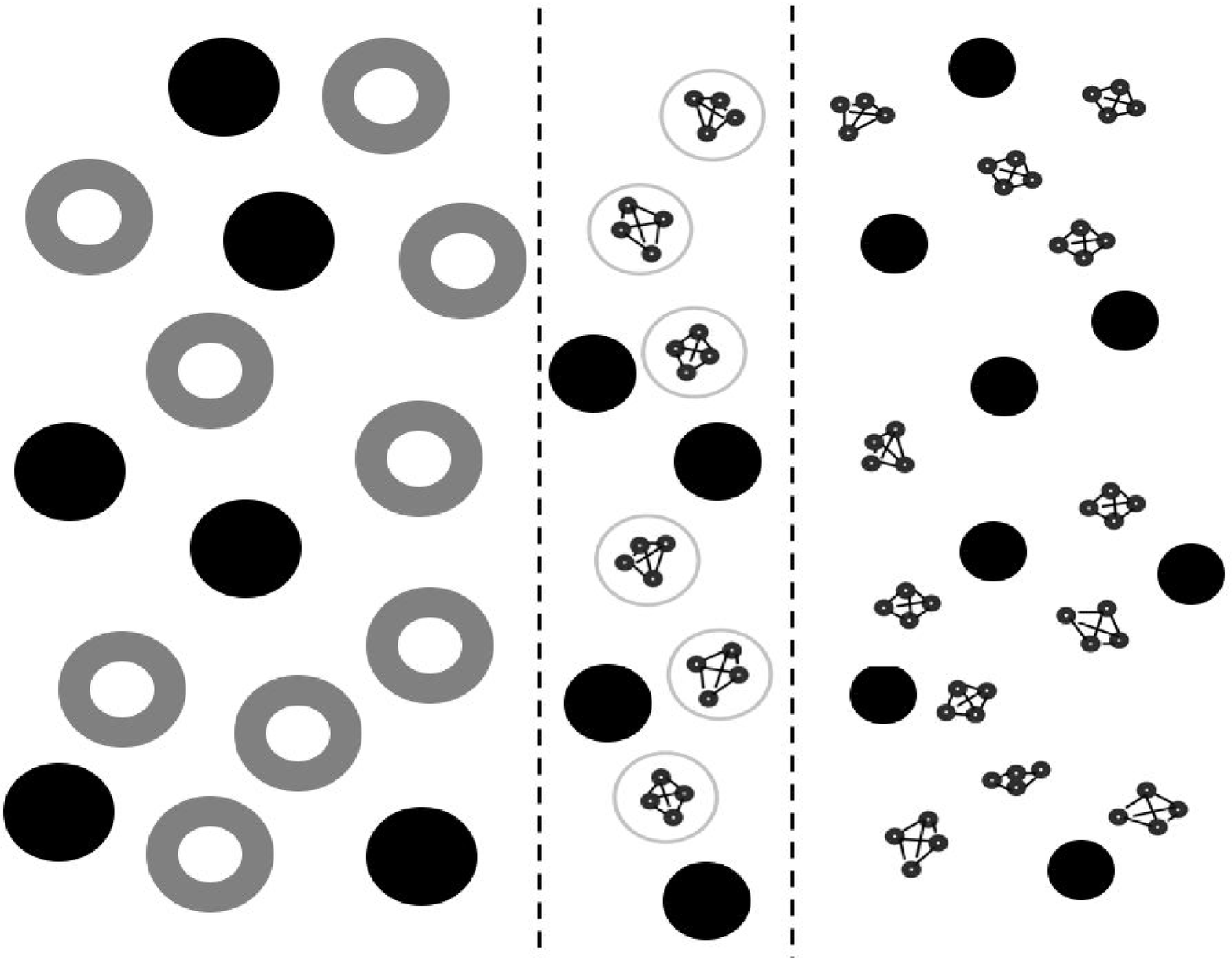}
\caption{Pictorial representation of the system. On the right the atomistic tetrahedral molecules solvating a spherical solute, in the middle a hybrid representation and on the left the coarse grained one. Notice that the solute, although in all cases one-sited, is characterized by different effective interactions and excluded volume in the different regions since it has to be consistent with the different resolutions of the solvent. \label{cartoon}}
\end{figure}

\begin{figure}[H]
\includegraphics[width=0.5\textwidth]{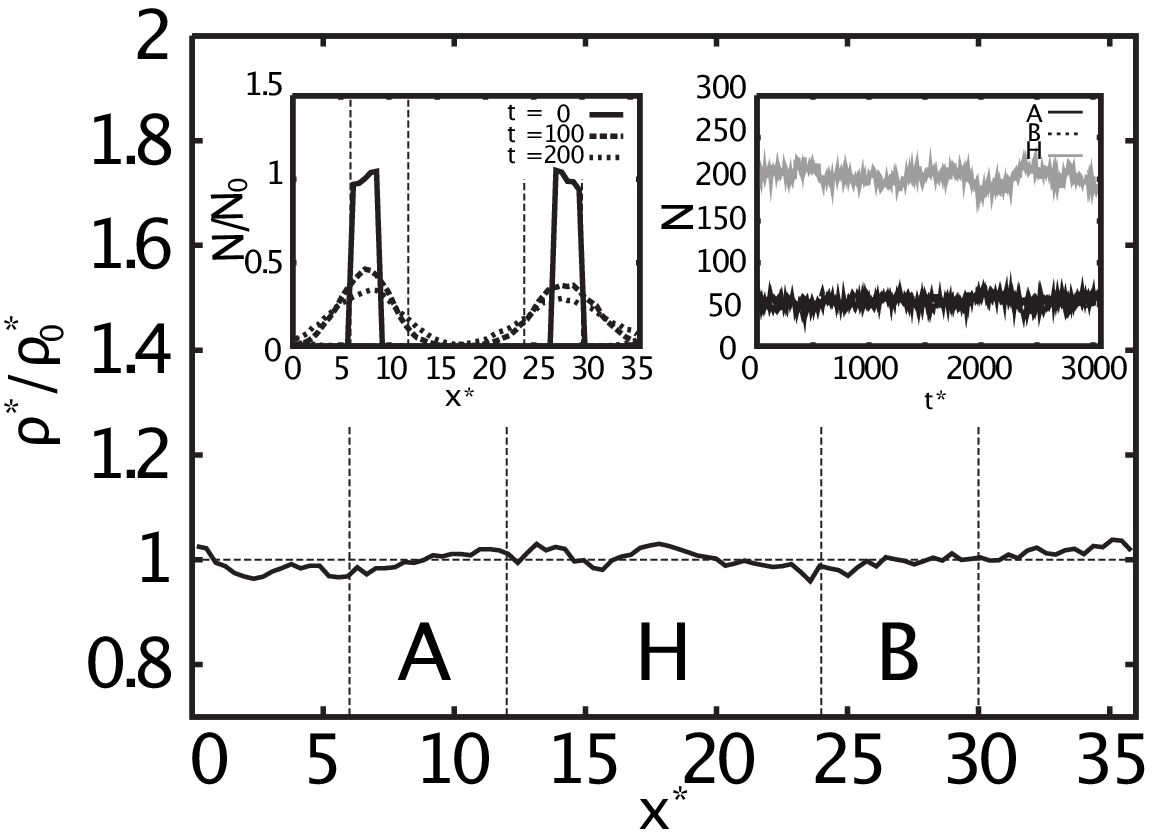}
\caption{Density profile of the solute across the box. The total number of molecules for the solute is 
311, while that of the solvent is 2174, corresponding to a solute concentration of 0.14. Inset, right panel, the total number of molecules in the three regions, as a function of time; $A$ atomistic, $H$ hybrid, $B$ coarse grained. 
Inset, left panel, diffusion profiles as a function of time of atomistic and coarse grained molecule across the box, the figure display a proper diffusion, assuring that there are no barriers across the system. The three plots show that the system is in a stationary (equilibrium) state. Note that the transition region is larger than the atomistic and the coarse-grained. This was made on purpose because the properties in this region are conceptually of major interest for the development of the current model with the thermodynamic force.\label{mono}}
\end{figure}

\begin{figure}[H]
\includegraphics[width=0.5\textwidth]{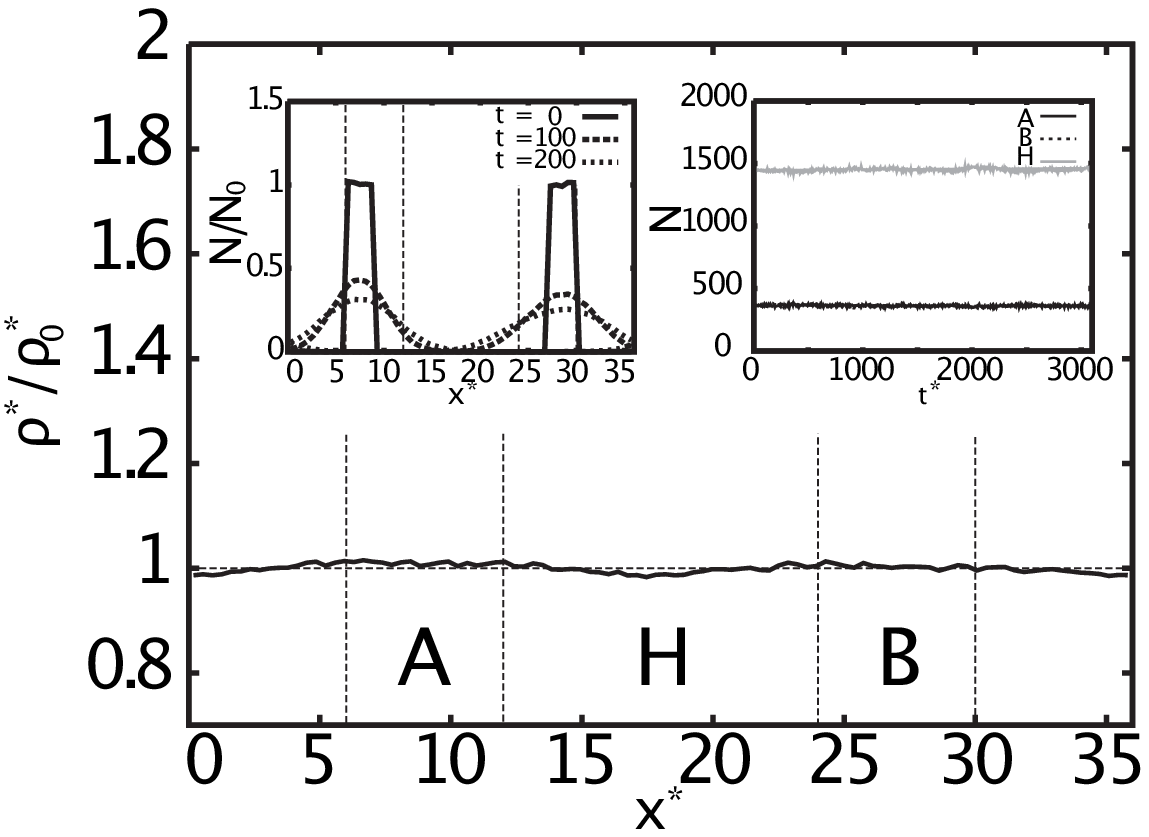}
\caption{As for Fig.\ref{mono}, now for the solvent. \label{solv}}
\end{figure}

We now apply the above developed concept to an "atomistic" liquid
of tetrahedral molecules which solvate another species of
spherical molecules (see the pictorial representation of the system, Fig. \ref{cartoon}).
 From
the atomistic simulation a coarse grained model of both solvent
and solute is derived and then the hybrid atomistic/coarse-grained
adaptive scheme is applied (for details of the model see note \cite{note5}). Doing this in the conventional way\cite{annurev},
leads to significant density variations throughout the system. As
for the one component system, we first calculate the chemical
potential for each species (solvent and solute) according to the
scheme previously shown. However, this is slightly more complex
than before, because we have two different molecules each with
its intrinsic chemical potential to which the contribution
originated by the mixing, both aspects then according to the
spatial resolution, must be added. With $c_{i}$ being
concentration of component $i$ we can write for the chemical potential of the solvent:
\begin{equation}
\mu^{mix}_{tetra}=\mu^{0}_{tetra}+kT\log[c_{tetra}]+f^{mix}_{int}(c_{tetra},c_{solute})
\label{mutetra}
\end{equation}
and equivalently for the solute:
\begin{equation}
\mu^{mix}_{solute}=\mu^{0}_{solute}+kT\log[c_{solute}]+g^{mix}_{int}(c_{tetra},c_{solute})
\label{muspheres}
\end{equation}
where $\mu^{0}$ is the chemical potential of the pure component at
the same density. $kT\log[c_{i}]$ is
the part coming from the entropy of mixing for the ideal noninteracting 
case for the solvent and solute, respectively.
$f^{mix}_{int}(c_{tetra},c_{solute})$ is the part originating from
the molecular interactions for the solvent and equivalently $g$
for the solute.  The functions $f$ and $g$ are unknown and
empirical expressions are given in literature (see e.g.
Ref.\cite{nieto,prausnitz}). However, here we have chosen to take a
more practical path to determine $f$ and $g$ which can
be easily implemented numerically and yet provides the internal
consistency the the whole algorithm of simulation. For this we
expand $f$ and $g$ as:
\begin{equation}
f^{mix}_{int}(c_{tetra},c_{solute})=\left[\frac{\partial f}
{\partial c_{tetra}}\right]_{ c_{tetra}^{0},c_{solute}^{0}}\cdot
c_{tetra} \label{exptetra}
\end{equation}
\begin{equation}
g^{mix}_{int}(c_{tetra},c_{solute})=\left[\frac{\partial g}
{\partial c_{solvent}}\right]_{ c_{tetra}^{0},c_{solute}^{0}}\cdot
c_{solvent} \label{expsph}
\end{equation}
here $c_{tetra}^{0},c_{solute}^{0}$ are some equilibrium
concentrations which as will be shown are not needed to be known
{\it a priori}. While all the other quantities entering the
chemical potentials are known, the question is how to obtain
$\left[\frac{\partial f}{\partial c_{tetra}}\right]_{
c_{tetra}^{0},c_{solute}^{0}}$ and $\left[\frac{\partial
g}{\partial c_{solvent}}\right]_{ c_{tetra}^{0},c_{solute}^{0}}$
as this is needed to obtain the overall thermodynamic force to be
applied to each component. To provide a practical approach we
first run an adaptive simulation of our system with a thermodynamic force without the terms corresponding to the mixing. Because in the thermodynamic force 
the terms of the mixing, at this stage, are neglected this simulation will
produce a non uniform density profile (or concentration profile)
in the transition region. Since we know from Eqs.\ref{mutetra},\ref{muspheres} and Eqs.\ref{exptetra},\ref{expsph} that the  terms coming from the mixing are functions of the density (concentration), we take the density profile obtained to determine the terms of mixing, tuning the unknown coefficients of $f$ and $g$ so that the corresponding (complete) thermodynamic force provides a flat profile.
By that, we numerically define the
unknown part of the chemical potential. As a test of consistency
we show that once these constants are fixed, the corresponding
thermodynamic force applied to different initial conditions keeps the profile flat and produces a stationary bidirectional flux of particles as for the one component system in Ref.\cite{jcp} (see Figs.\ref{mono},\ref{solv}). This is a
practical, yet powerful way to determine in general the chemical
potential profile of a mixture, consistent with the basic
thermodynamical principles of the adaptive representation.
\section{Conclusion/Outlook}
Our simulation results show how to set up a consistent framework for an
adaptive resolution simulation of solutions and mixtures. By the
introduction of the concept of a thermodynamic force, based on a
locally variable chemical potential, typical artifacts of such a
concurrent simulations with variable resolution can be avoided.
The method allows for a free exchange of molecules between
different regimes and the molecules adapt their very
representation according to the region they are in. The approach
easily can be extended to more complicated situations, such as
systems of even more components. For practical implementations one
actually does not have to resort explicitly to the formal
derivation via the chemical potential. As illustrated for the case
of a mixture, the density profiles in the transition regime can be
flattened by a thermodynamic force obtained by a simple numerical tuning. 
Of course a more formal way to determine the
force would end up with the same result.  The present approach
however is even more general than discussed so far. Usually one
deals with a well defined system which one wants to study in
different regions of space by different resolutions. This allows
to zoom in within a molecular simulation and to study regions of
special interest in more detail. The present ansatz on the other hand 
can be extended to a much wider class of problems. There is absolutely no
reason to restrict the method to the case of $\mu= 0$ in the
pure atomistic and coarse grained region. In principle on can by
such a method couple systems with (almost) arbitrary differences
and keeps them in equilibrium with each
other. Though this might look a bit unexpected, this allows for
instance to introduce concepts of open systems or grand canonical
molecular dynamics simulations. This will also be of special
interest when it comes to non-equilibrium situations like the
change of concentrations of one species in the surrounding etc.

\begin{acknowledgments}
The authors acknowledge the constructive discussions with B.D\"{u}nweg, C.Peter, D.Andrienko and R.Delgado-Buscalioni. This work was partially supported by the  DAAD-Conicyt grant provided to S.P. 
\end{acknowledgments}


\begin{thebibliography}{8}
\bibitem{note2}
Alternatively, one
also could use the pressures and, for mixtures, partial pressures
for a similar scheme. Because pressure only requires forces, it
is well defined within the whole simulation box.  We here however
use the chemical potentials, since this allows us more easily to
formulate the setup in a more general way.
\bibitem{note3}
In practice,
this is exact only for a vanishing slope of $w(x)$, since in the
actual simulation in the hybrid regime the molecules interact with
other molecules of slightly different representation. This
however, as will be seen below, is a rather good approximation.
\bibitem{note4}
Actually, this method could also be used to introduce
a density gradient on purpose and thus to couple rather different
systems.
\bibitem{note}
The kinetic energy of a given hybrid molecule can be written
as\cite{jpa}:
$K=K_{trans}+K_{intra}=K_{cg}+K_{intra}=K_{cg}+w
K_{intra}+(1-w)K_{intra}=K_{cg}+K_{fractal/explicit}+K_{fractal/internal}$,
where $K_{cg}$ is the kinetic energy of the coarse-grained
molecule, $K_{intra}$ is the intramolecular kinetic energy
corresponding to molecular rotational and vibrational DOFs (these
DOFs are switched on/off while changing the resolution),
$K_{fractal/explicit}$ is the kinetic energy associated with the
fractal rotational+vibrational DOFs that are explicitely
considered in the simulation, $K_{fractal/internal}$ is the total
internal kinetic energy associated with those fractal DOFs that
are not explicitely considered in our treatment, and $w$ is the
degree of fractionality (the degree of how much a particular DOF is
switched on). During the adaptive resolution simulation 
the explicit DOFs of the hybrid molecules, i.e.,
rotational + vibrational DOFs, are also equilibrated. 
The explicit DOFs of a given hybrid molecule are fully present in the simulation regardless of the hybrid molecules' position in the transition regime but they contribute differently to the statistical averages (depending on $w$). The switching on/off DOFs has to be understood in the statistical sense, i.e., how much of a particular DOF contribution is explicitly considered in the statistical-mechanical averages.  Therefore, the thermostat has to supply or remove the same amount of kinetic energy to/from the explicit DOFs of the hybrid molecules in the transition regime as it does in the case of the fully explicit molecules in the all-atom regime in order to keep the system at given temperature. 
The $w$ part of this energy contributes to the $K_{fractal/explicit}$ while the $1-w$ part is ascribed to $K_{fractal/internal}$. The latter represents a part of the latent heat associated with the change of the resolution.
\bibitem{note5}
We have developed the coarse-grained potentials that reproduce the thermodynamic state point and basic structure of two atomistic mixtures: one, at densities of 0.174 for the solvent and 0.001 for the solute, in Lennard-Jones reduced units, and a more concentrated system with densities of 0.151 and 0.022 respectively. In both cases the pressure matches the value obtained from the pure tetrahedral liquid previously studied.
The interactions were parameterized using the Iterative Boltzmann method as in Ref.\cite{jcp}. By these means, we obtained a first guess for the coarse-grained potential between the two species for the most diluted mixture.
Next we proceeded to reparametrize the solvent-solvent interaction on the most concentrated system and further tune the potential between both species to conclude with the refinement of the force field between the solute particles until the radial distribution functions between solvent, solute and both kinds of particles are in good agreement with the atomistic simulations.
In addition, the set of potentials obtained in this way reproduces accurately the pressure for the highest concentration and below
\end{thebibliography}
\end{document}